\documentclass[%
reprint,
superscriptaddress,
amsmath,amssymb,
aps,
]{revtex4-2}

\usepackage{graphicx}
\usepackage{dcolumn}
\usepackage{bm}
\usepackage{xr}
\usepackage{float}

\begin{document}
	
	
	\title{Tailoring 3D Speckle Statistics}
	
	\author{SeungYun Han}
	\affiliation{%
		Department of Applied Physics, Yale University, New Haven, Connecticut 06520, USA
	}
	\author{Nicholas Bender}%
	\affiliation{%
		School of Applied and Engineering Physics, Cornell University, Ithaca, New York 14850, USA
	}
	\author{Hui Cao}%
	\email{hui.cao@yale.edu}
	\affiliation{%
		Department of Applied Physics, Yale University, New Haven, Connecticut 06520, USA
	}%

	\date{\today}
	
	\begin{abstract} 
		
		We experimentally generate three-dimensional speckles with customized intensity statistics. By modulating the phase front of a laser beam, far-field speckle patterns maintain the designed intensity probability density function while evolving to different spatial patterns upon axial propagation. We also create speckles with distinct intensity statistics at multiple designated planes. These results open many new possibilities, such as designing volumetric speckle statistics for three-dimensional imaging, sensing, trapping, and manipulation.
		
	\end{abstract}
	
	\maketitle
	
	
	
    Speckles occur in many types of waves: electromagnetic, acoustic, and matter. They are produced by the interference of numerous partial-waves with uncorrelated phases. Characterized by a random granular pattern of diffraction limited grains, fully-developed speckles generally satisfy Rayleigh statistics \cite{goodman2007speckle, dainty2013laser}. Recent studies, however, have shown that fully-developed speckle patterns can be customized to possess circular non-Gaussian statistics in a single plane: i.e. an arbitrary probability density function and/or long-range correlations \cite{bromberg2014generating, bender2018customizing, bender2019introducing, bender2019creating, bender2021circumventing}. In these works, the customized properties erode away with axial propagation: eventually, the speckles revert back to Rayleigh statistics. Whether it is possible to create axially-varying volumetric-speckles with tailored statistics remains an open question. The major hindrance to creating such volumetric customized speckles is the fact that the field profiles at different axial planes are related: specifically, the field profile at one axial plane determines the fields at \textit{all} planes after it.

    The recent interest in customizing properties of speckle patterns~\cite{bromberg2014generating, bender2018customizing, bender2019introducing, bender2019creating, bender2021circumventing, dogariu2015electromagnetic, fischer2015light, amaral2015tailoring, kondakci2016sub, li2016generation, di2016tailoring, di2018hyperuniformity, devaud2021speckle, liu2021generation} stems, in large part, from their numerous practical applications. For example, optical speckles have been widely employed in structured-illumination imaging \cite{lim2008wide, singh2017exploiting, vigoren2018optical, min2013fluorescent, lim2011optically}, and sensing applications \cite{anand2007wavefront, kim2016remote, berto2017, luo2021super} such as digital holography \cite{bernet2011lensless, baek2020speckle}, ghost imaging \cite{zhang2015ghost, phillips2016non, liu2020spectral, nie2021noise}, computational imaging \cite{mudry2012structured, yilmaz2015speckle, pascucci2019compressive, yeh2019speckle}, dynamic speckle illumination microscopy \cite{ventalon2005quasi, choi2022wide}, Fourier ptychography \cite{zhang2019near}, and photoacoustic imaging \cite{chaigne2016super}. While Rayleigh speckles are commonly used, super-Rayleigh and other types of customized speckles improve high-order correlation-based imaging techniques, providing better visibility and a higher signal-to-noise ratio \cite{oh2013sub, kuplicki2016high, li2021sub}. They can also boost super-resolution imaging techniques, obtaining higher resolutions than Rayleigh speckles \cite{zhang2016high, liu2019label}. Specially designing the intensity statistics of speckles used in parallelized nonlinear pattern-illumination microscopy enables a three-times higher spatial resolution than provided by the optical diffraction limit \cite{bender2021circumventing}. However, the axial resolution has not yet been improved. Tailoring the statistics of speckles along the axial direction can provide a better axial resolution in three-dimensional (3D) imaging \cite{pascucci2019compressive}. This may be done by creating volumetric speckles, which axially decorrelate yet retain the same statistics, or even by creating speckles with distinct intensity statistics upon axial propagation.	Moreover, the optical potentials created by 3D customized speckles can provide an effective platform for studying 3D transport and localization of cold atoms and active media in random potentials \cite{douglass2012superdiffusion, levi2012hyper, nagler2022ultracold}.  They may also be used to manipulate, sort, and order microparticles in 3D \cite{dholakia2011shaping, volpe2014speckle, volpe2014brownian, nunes2020ordering}. 
    %
	\begin{figure*}[ht]
	 \centering
		\includegraphics[width=.9\textwidth]{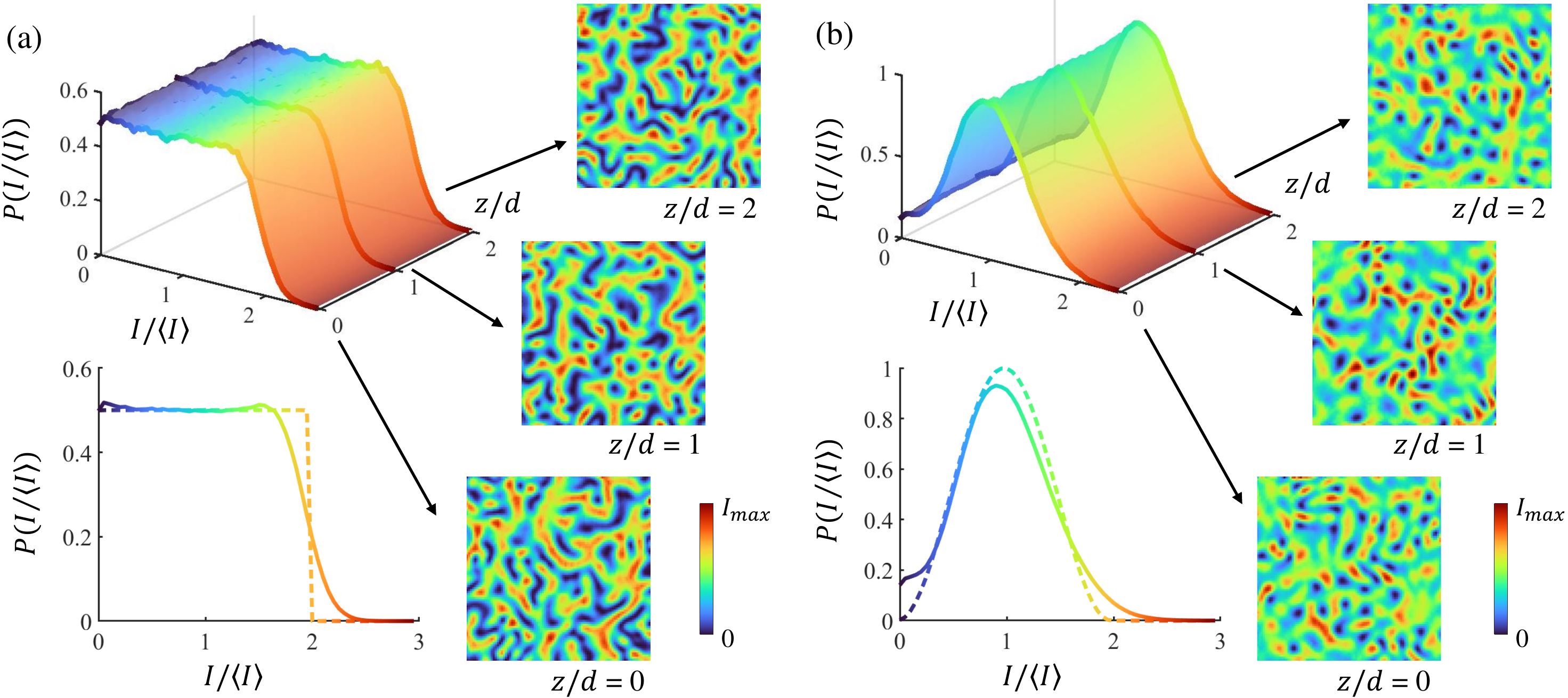}
		\caption{\label{volume} {\bf Experimental realizations of volumetric speckles with customized intensity PDFs}. $P(I/ \langle I \rangle)$ is uniform within the range $0 \leq I/ \langle I \rangle \leq 2$ in (a), and peaked at $I/ \langle I \rangle = 1$ in (b). Top left graph in (a,b) shows axial-invariant intensity PDF, bottom left graph is the intensity PDF obtained within two Rayleigh ranges. Speckle patterns evolve axially as shown in the right for axial locations $z/d=0, 1, 2$. The Rayleigh range $d$ = 9.3 cm. Each speckle image corresponds to $90 \times 90$ camera pixels ($504\times504$ ${\rm \mu m}^2$). Intensity PDFs are obtained by averaging over 30 speckle realizations. }
	\end{figure*}
	%
	\newline \indent In this Letter, we experimentally demonstrate a method of customizing the 3D intensity statistics of speckles. With it, we can control the intensity probability density functions (PDFs) of speckle patterns on multiple axial planes. By appropriately manipulating the phase front of a laser beam, with a spatial light modulator, the far-field speckles can maintain a desired intensity PDF while propagating and evolving into distinct patterns over two Rayleigh ranges (equal to the longitudinal speckle grain size). Our results show that the high-order correlations of partial waves, which are determined by the intensity PDF, can be controlled over an extended axial range despite the addition of quadratic phase. Our method requires phase-only modulation, which minimizes power loss relative to the amplitude \& phase modulation used for generating non-diffracting speckles \cite{liu2021generation}. Our work opens the door toward tailoring volumetric light statistics for 3D imaging and sensing applications.
	%
	%
    \newline \indent Our experimental setup is simple and easy to implement \cite{bender2019creating}. It consists of a laser (wavelength $\lambda=638$ nm), a phase-only spatial light modulator (SLM, Hamamatsu LCOS-SIM), an optical lens (focal length = 20 cm) and a CCD camera (Allied Vision Prosilica GC660). The SLM is placed at the front focal plane of the lens, and is fully illuminated by the linearly-polarized laser beam. The light reflected from the SLM passes through the lens, and the first-order diffraction is filtered by an iris to remove any unmodulated light. The filtered light then reaches a camera, which is mounted on a motorized translation stage that can axially move over two Rayleigh ranges, starting from the back focal plane of the lens. When a random phase pattern is displayed on the SLM, a Rayleigh speckle pattern is generated at the back focal plane of the lens ($z$ = 0). It evolves upon axial propagation and becomes decorrelated after one Rayleigh range $\Delta z = d$ = 9.3 cm (given by the width of axial correlation function of speckle intensity). To a good approximation, the field incident on the camera (at $z$ = 0) is a Fourier transform of the field reflected off the SLM. To be more precise and general, we measure the field-transmission matrix (T-matrix) between them, to account for experimental artifacts such as lens aberrations and optical misalignment.

    Two examples of experimentally measured 3D-speckles, with distinctly customized intensity statistic, are presented in Fig.~\ref{volume}. An example of volumetric speckles with an intensity PDF tailored to be flat, $P(I/ \langle I \rangle)= 0.5$ over a predefined range of intensity values $0 \leq I/ \langle I \rangle \leq 2$, is shown in Fig.~\ref{volume} (a), where $\langle I \rangle$ denotes the mean intensity. The 2D speckle pattern in the Fourier-plane $(z=0)$ axially evolves and eventually decorrelates with its original spatial profile at $ z = d$ (see Fig. S1 in Supplementary Materials). Even though the speckle pattern decorrelates, the intensity PDF remains flat over two Rayleigh ranges $0 \leq z \leq 2d$. Similarly, Fig.~\ref{volume}(b) is an example of volumetric speckles with a unimodal intensity PDF. Again, the PDF is axially invariant despite the axial correlation of the speckles: therefore, the customized intensity PDF describes volumetric speckle statistics. That it is possible to design volumetric speckle fields, however, is quite counter-intuitive.

	The major challenge we combat when customizing the statistics of 3D speckles, like those shown in Fig.~\ref{volume}, is the fact that the 2D fields on different axial planes are related. Specifically, the field profile at one axial plane will determine the fields at all subsequent propagation distances along $z$. This propagation relation is not a concern when optimizing speckle statistics on a single 2D-plane. In 2D, a speckle pattern can be customized via the following process~\cite{bender2018customizing}. First, a target speckle intensity pattern which obeys the desired intensity PDF is numerically generated by transforming a Rayleigh speckle pattern. Second, the SLM phase pattern for generating the target intensity profile is found using a T-matrix-based nonlinear-optimization algorithm. This technique cannot be directly used for 3D speckle customization since, upon application of the local intensity transformation to the different axial planes, the resulting fields no longer self-consistently satisfy the propagation relation necessary for the fields to be physically possible.
	
	\begin{figure}[htb]
		\includegraphics[width=8.5cm]{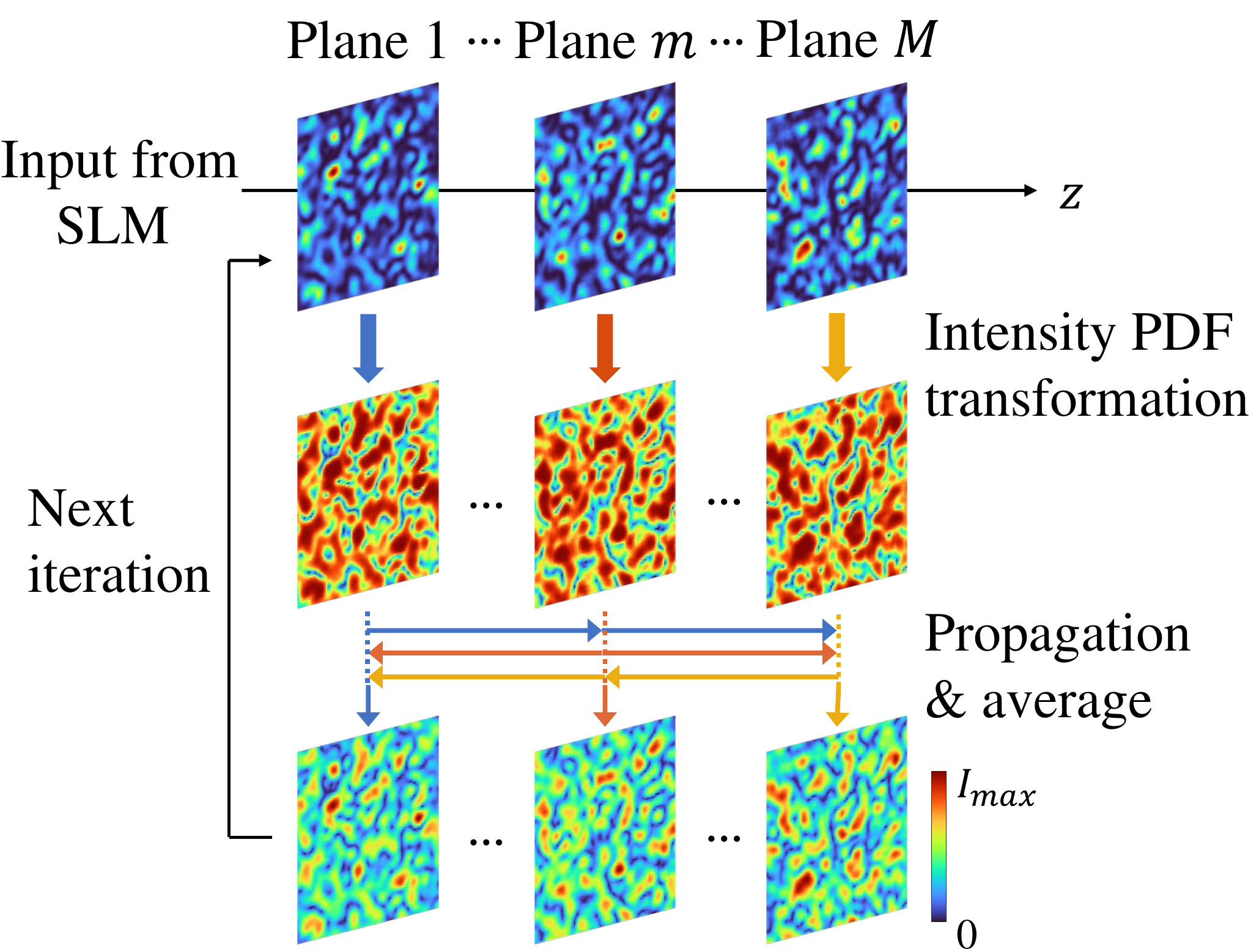}
		\caption{\label{algorithm} {\bf Numerical algorithm for tailoring 3D speckle statistics.} A schematic illustration of customizing the speckle intensity PDFs across $M$ axial planes. Starting with Rayleigh speckles generated by a random phase pattern on the SLM, a local intensity transformation is applied independently to each plane to obtain the target PDF. Modified fields in one plane propagate to all other planes and then are averaged. This process is iterated until the fields in all planes satisfy the propagation relation and the target intensity PDF.}
	\end{figure}
	
    The algorithm we use to overcome the propagation-relation challenge is illustrated in Fig.~\ref{algorithm} and can be described as follows. For simplicity, we attempt to customize the volumetric statistics over two Rayleigh ranges ($\Delta z = 2 d$). Since the speckles only change slightly over a propagation distance $\Delta z = 0.2 d$, we optimize the speckles on 11 equally-spaced axial-planes from $z$ = 0 to $2d$. With random phase modulation of a $32 \times 32$ macropixel array on the SLM, Rayleigh speckle patterns in the 11 planes are numerically calculated with the measured T-matrix. To encode the desired intensity PDF in these planes, a local intensity transformation is applied independently to the spatial intensity-pattern in each plane, keeping the spatial phase-pattern unchanged. This process breaks the propagation relation between the planes, which is then partially recovered by a procedure of propagation and averaging. Namely, to correct the field in one plane, the fields in all other planes are numerically propagated to this plane and then averaged. The same process is repeated for the remaining 10 planes. The updated fields are transformed to the SLM plane to remove amplitude modulation, and then transformed back. The resulting fields are used for the next iteration of intensity transformation, until the fields on all planes satisfy the propagation relation and the target intensity PDF. Typically, the iteration converges after 1,000 cycles, providing the phase-only modulation pattern on the SLM. Repetition of this procedure with different initial random phase patterns on the SLM creates a set of independent 3D speckle patterns that possesses the same intensity PDFs. Note that, the axial spacing of the measured planes $\Delta z = 0.05d$ in Fig.\ref{volume} is much smaller than that of 11 planes $\Delta z = 0.2d$ in the numerical algorithm, confirming that it is sufficient for ``volumetric'' control of speckle statistics by making the intensity PDF identical at axial planes separated by $0.2d$.  We verify that the speckles generated with this technique are fully developed \cite{bender2019creating}, by calculating the joint complex-field PDF shown in Fig. S2 of Supplementary Materials. Its circularity reveals that the speckle fields have isotropic phase distribution over $[0, 2\pi)$, thus they are fully developed.

	\begin{figure}[htb]
		\includegraphics[width=8.5cm]{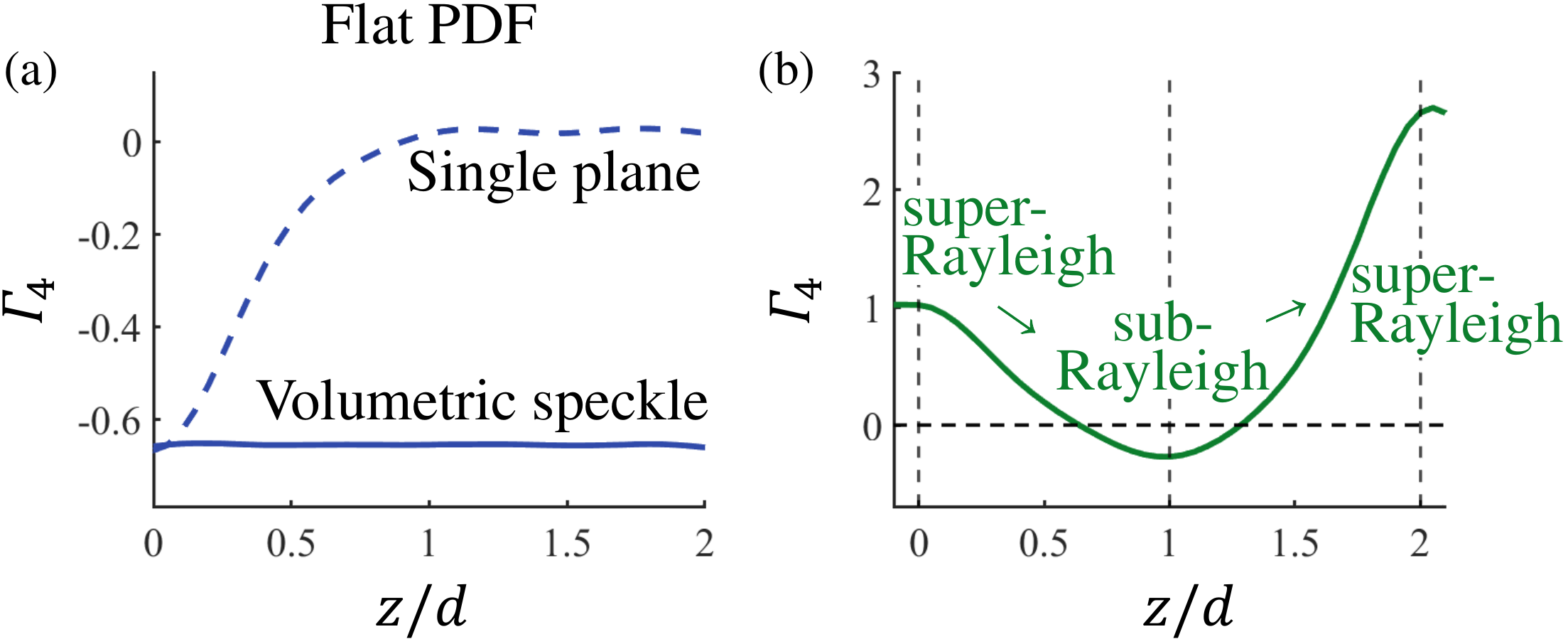}
		\caption{\label{Gamma4} {\bf High-order correlations of Fourier fields}. (a) $\Gamma_4(z) = C^2(z) - 1$ for volumetric speckles in Fig.~\ref{volume}(a) with flat intensity PDF (blue solid curve) stays nearly constant with $z$. For comparison, $\Gamma_4(z)$ returns to 0 (blue dashed curve) within one Rayleigh range $d$, when the intensity PDF is flat only at a single plane ($z=0$). (b) Axial variation of $\Gamma_4$ when speckle intensity PDF transforms from super-Rayleigh at $z/d =0$ to sub-Rayleigh at $z/d =1$ and back to super-Rayleigh at $z/d =2$ in Fig.~\ref{3plane}(b). $\Gamma_4$ changes sign twice, since its value is positive for super-Rayleigh and negative for sub-Rayleigh.  $\Gamma_4$ values in (a,b) are obtained by averaging over 30 speckle realizations.}
	\end{figure}
	
	Through the customization, our algorithm encodes high-order correlations into the speckle fields. This can be seen via the speckle intensity contrast. The intensity contrast $C$ of the customized speckles is determined by the target PDF, and deviates from 1 for Rayleigh speckles. As derived in Ref.\cite{bromberg2014generating}, $C^2(z) \simeq 1 + \Gamma_4(z)$, and
	\begin{equation} 
	\Gamma_4(z) \propto \sum \tilde{E}({\bf k}_1,z) \, \tilde{E}({\bf k}_2, z) \, \tilde{E}({\bf k}_3, z) \,  \tilde{E}({\bf k}_1-{\bf k}_2-{\bf k}_3, z), 
	\label{Gamma4Eq}
	\end{equation}
	where $\tilde{E}({\bf k}, z)$ is transverse Fourier transform of speckle field $E({\bf r}, z)$, ${\bf k}$ is the transverse spatial vector, and the summation in Eq.~\ref{Gamma4Eq} is over ${\bf k}_1$, ${\bf k}_2$ and ${\bf k}_3$ with the constraint ${{\bf k}_1 \neq{\bf k}_2 \neq {\bf k}_3}$. Thus, non-Rayleigh speckle statistics $C \neq 1$ at $z$ = 0 results from high-order correlations $\Gamma_4$ encoded into the Fourier fields $\tilde{E}({\bf k}, z)$. If only the speckle statistics on one plane are customized, $\Gamma_4$ will vanish upon axial propagation from it, and $C$ will return to 1 within one Rayleigh range [dashed line in Fig. \ref{Gamma4} (a)]. The axial propagation corresponds to adding a quadratic phase-front to the optimized Fourier field profile, which tends to destroy the high-order correlation. For the volumetric customized-speckles, the high-order correlations among the Fourier fields $\tilde{E}({\bf k}, z)$ are maintained over extended axial propagation. This is shown by the solid line in Fig.~\ref{Gamma4}(a), where $\Gamma_4$ stays at a non-zero and constant value across two Rayleigh ranges.  
	
	\begin{figure*}[ht]
		\includegraphics[width=16cm]{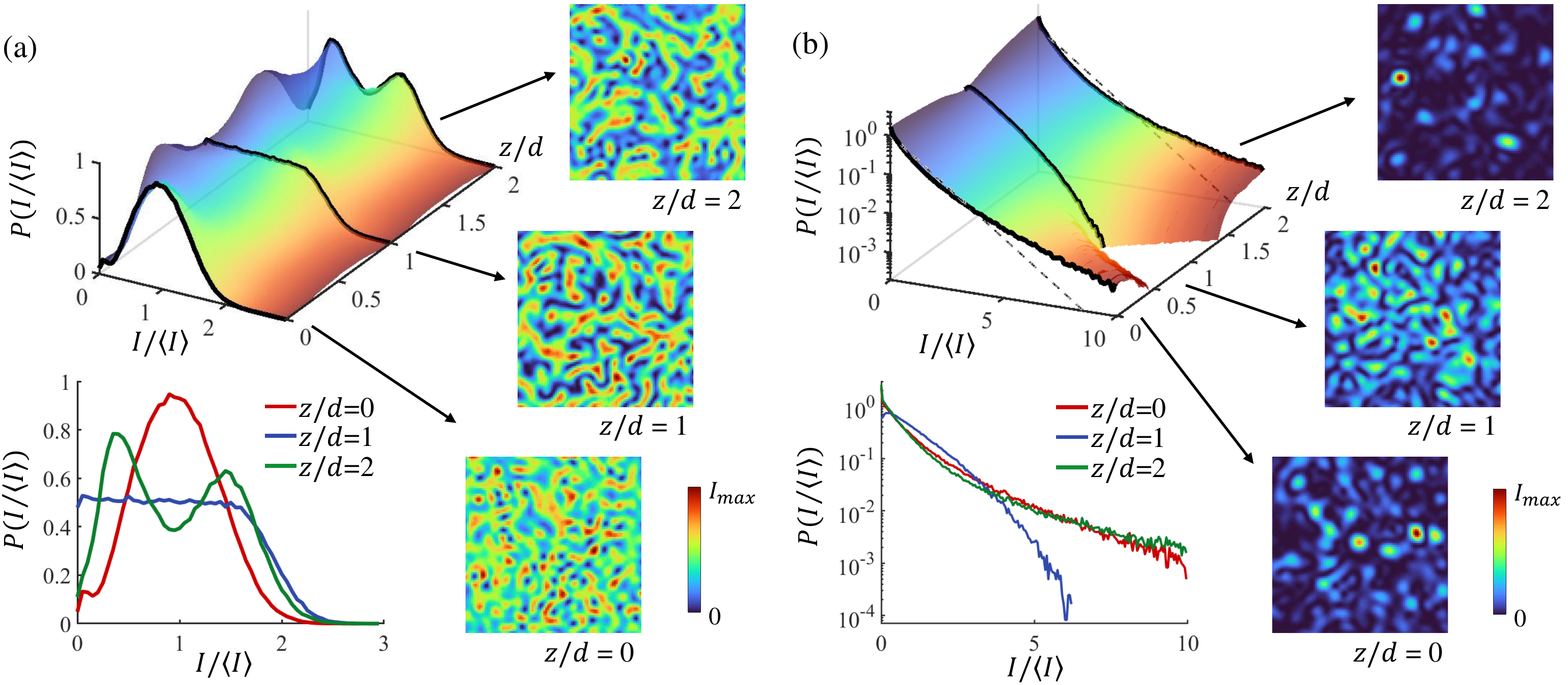}
		\caption{\label{3plane} {\bf Experimental realizations of distinct speckle statistics on three planes}. Speckle intensity PDFs at three axial planes, $z/d=0, 1, 2$, are unimodal, flat, bimodal in (a); super-Rayleigh, sub-Rayleigh, super-Rayleigh in (b). Top left graph in (a,b) shows an axially-evolving intensity PDF, bottom left graph is the intensity PDF at $z/d=0, 1, 2$. Speckle patterns at the three designated planes are shown on the right. All parameters are identical to those in Fig.~\ref{volume}.}
	\end{figure*}
	
	Until now, we have focused on generating speckles with a uniform volumetric PDF. Nothing precludes us, however, from using the same technique for a different goal: generating speckles with axially varying PDFs. Here, we demonstrate that our method is also able to create speckles with distinct statistics on different axial planes. As an example, we select three planes separated by one Rayleigh range, $z = 0, d, 2d$. As shown in Fig.~\ref{3plane}(a), speckle intensity PDFs for the three planes are designated to be unimodal, flat, and bimodal. We apply the same algorithm for volumetric speckles to the three planes, and obtain the SLM phase pattern. Experimentally we record the speckle patterns with the axial range of $0\leq z \leq 2d$ and plot the intensity PDF for each plane.  Fig.~\ref{3plane}(a) shows the measured PDF evolving from unimodal at $z = 0$ to flat at $z=d$, and to bimodal at $z= 2d$. The speckle fields on these planes satisfy the propagation relation. Thus, the intensity PDFs on different planes can be tailored simultaneously and independently for axial-evolving speckle patterns. The separation between the designated planes may vary, but should exceed the minimum distance ($0.2d$) to allow the intensity PDF to evolve to a different one. 	
	A second example is given in Fig.~\ref{3plane}(b), where the speckles evolve from super-Rayleigh at $z =0$ to sub-Rayleigh at $z=d$ and back to super-Rayleigh at $z=2d$. Here, the super- and sub-Rayleigh refer to the intensity PDF that decays faster and slower than the exponential decay of Rayleigh PDF, respectively \cite{bromberg2014generating}. $\Gamma_4$ is positive for super-Rayleigh, and negative for sub-Rayleigh. Figure~\ref{Gamma4}(b) shows the axial variation of $\Gamma_4$ over two Rayleigh range. Starting from a positive value at $z=0$, it drops and crosses zero to reach a negative value at $z=d$, then it rises to cross zero and become positive again at $z=2d$. Therefore, a single SLM phase pattern can encode distinct high-order correlations among Fourier components of speckle fields at varying axial locations.
	\newline \indent In conclusion, we have developed an efficient method for customizing 3D speckle intensity statistics and experimentally demonstrated it using a phase-only SLM. Our method of 3D customization of speckle statistics is compatible with a broad range of experimental setups. It paves the way to new directions in both fundamental research (transport and localization of cold atoms and colloidal particles in 3D random optical potentials with tailored statistics) and applied research (3D speckle-based imaging, holography sensing, trapping and manipulation). For example, axial evolution from super-Rayleigh to sub-Rayleigh speckles leads to a varying intensity contrast, which is useful for axial resolution or 3D optical control of atoms/particles. In dynamic speckle illumination microscopy and HiLo microscopy \cite{ventalon2005quasi, mazzaferri2011analyzing}, the speckle contrast is an important factor that determines the axial sectioning ability, and a judicious manipulation of speckle contrast may improve the axial resolution. While the axial range has been within two Rayleigh ranges in our demonstration, this can be extended as long as there are sufficient degrees of control by the SLM. Therefore, the volumetric control of speckle statistics provides a powerful tool that expands the customized speckles to a full 3D regime. 
	\bigskip
	
	This work is supported by the US Office of Naval Research (ONR) under Grant Nos. N00014-21-1-2026 and N00014-20-1-2197.
	
	
	\bibliography{main}
	
\end{document}



\title{Supplementary Material: Tailoring 3D Speckle Statistics}

\author{SeungYun Han}
 \affiliation{%
 Department of Applied Physics, Yale University, New Haven, Connecticut 06520, USA
}
\author{Nicholas Bender}%
  \affiliation{%
 School of Applied and Engineering Physics, Cornell University, Ithaca, New York 14850, USA
}
\author{Hui Cao}%
 \email{hui.cao@yale.edu}
\affiliation{%
 Department of Applied Physics, Yale University, New Haven, Connecticut 06520, USA
}%
%


\begin{abstract}
		
	This document provides supplementary information to ``Tailoring 3D Speckle Statistics''. We show the axial decorrelation of volumetric speckles with tailored intensity statistics in section I. Section II presents joint complex-field probability density functions of 3D customized speckles.
		
\end{abstract}

\maketitle

\section{Axial decorrelation of volumetric speckles}

\begin{figure}[b]
\renewcommand{\figurename}{Figure S}
\includegraphics[width=.43\textwidth]{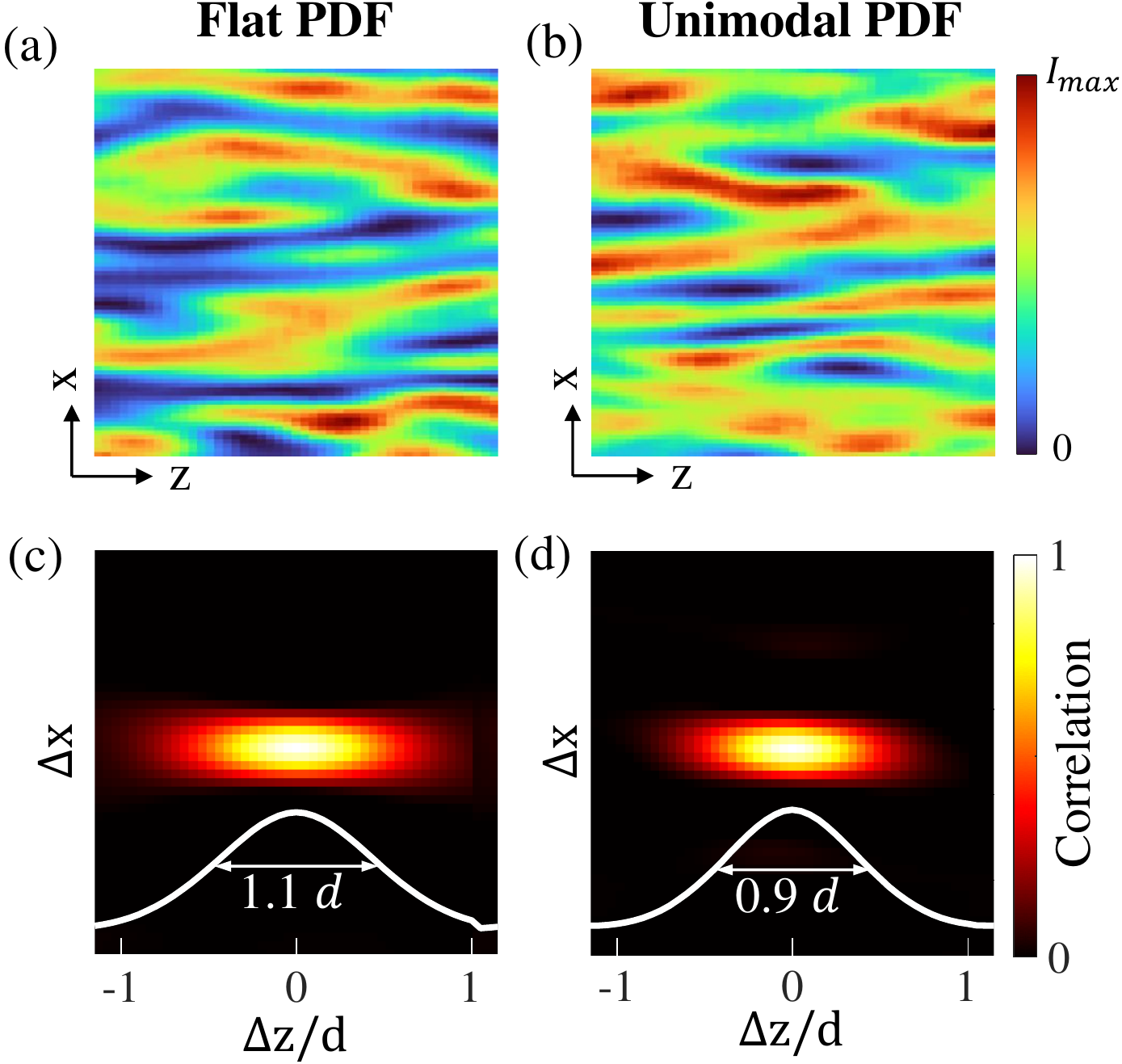}
\caption{\label{axial} (color online). {\bf Axial decorrelation of customized volumetric speckles}.  (a,b) $x$-$z$ cross-sections of speckles with flat and unimodal intensity PDFs, respectively. The intensity profile evolves axially, while the intensity PDF remains the same.  The cross-section has the transverse width 504 $\mu $m and axial length 18.6 mm. The Rayleigh range is $d$ = 9.3 mm. 
	(c,d) Spatial intensity correlation function $C_I(\Delta x, \Delta z)$ is computed from (a,b). The FWHM of $C_I(\Delta x, \Delta z)$ in $\Delta z$ (white curve) gives the axial correlation length (as marked), which is close to that of Rayleigh speckles ($d$). }
\end{figure}

Our method allows for designing the speckle intensity PDF within a volume. While the intensity PDF remains the same, the speckle pattern itself evolves axially. Figures S\ref{axial}(a, b) are the $x$-$z$ cross-sections of two example speckle patterns with flat and unimodal intensity PDFs, shown in Figs. 1(a,b). The intensity profile $I(x, z)$ changes gradually along the axial direction. We compute the intensity correlation function: $C_I(\Delta x, \Delta z) \equiv \langle I(x, z) \, I(x+ \Delta x, z + \Delta z) \rangle_{x,z} - 1$, after normalization $\int I(x, z) \, dx = 1$. The axial correlation length is defined as the full-width-at-half-maximum (FWHM) of $C_I(\Delta x, \Delta z)$ in $\Delta z$.  As shown in Fig. S\ref{axial}(c,d), the axial correlation lengths for both customized volumetric speckles are very close to the Rayleigh range $d$, which is defined as the FWHM of $C_I(\Delta x, \Delta z)$ for Rayleigh speckles. Therefore, the customized speckles with axially-invariant PDFs have axial intensity-decorrelations similar to Rayleigh speckles.

\section{Joint complex-field PDFs of customized speckles}
\begin{figure}[b!]
	\renewcommand{\figurename}{Figure S}
	\includegraphics[width=.45\textwidth]{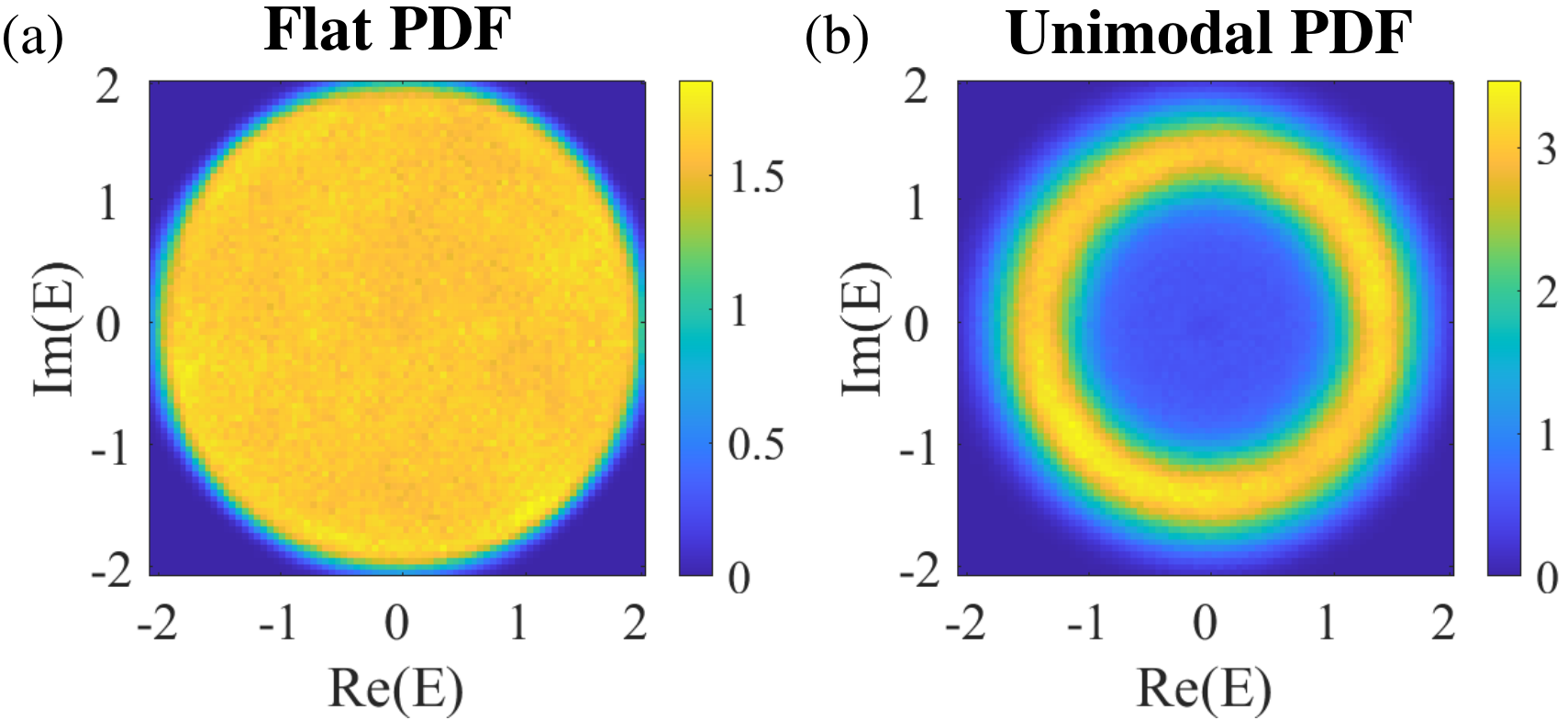}
	\caption{\label{joint1} (color online). {\bf Joint complex-field PDF of volumetric speckles with axial-invariant intensity PDF}. All parameters are identical to those in Fig. 1. The axial-invariant intensity PDF is flat in (a), and unimodal in (b). In both cases, the phase of the speckle field is uniformly distributed between $[0, 2\pi)$, indicating the customized speckles are fully developed. Joint complex-field PDFs are obtained from speckle fields over two Rayleigh ranges and in 30 speckle realizations.}
\end{figure}

\begin{figure*}[th!]
	\renewcommand{\figurename}{Figure S}
	\includegraphics[width=.9\textwidth]{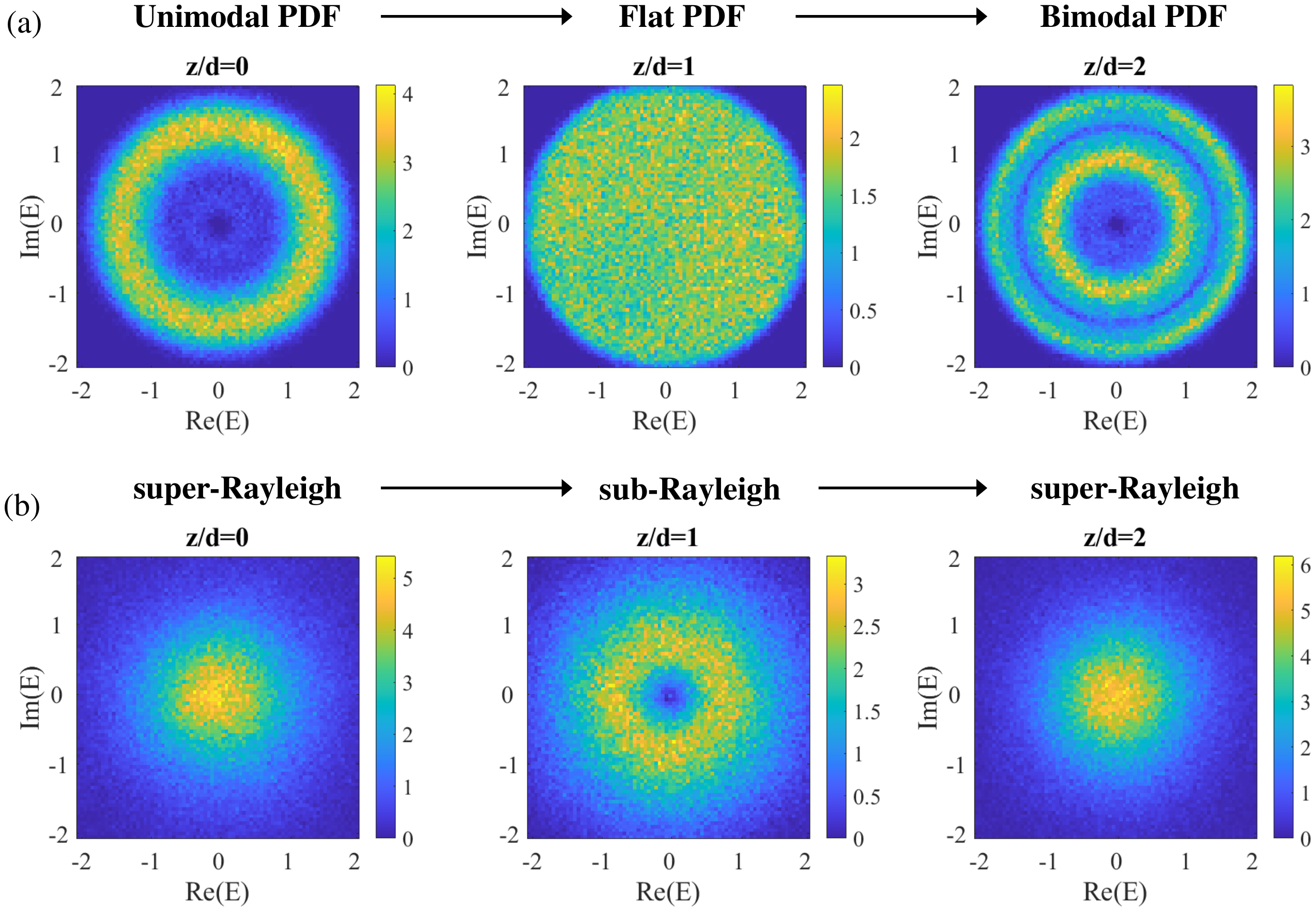}
	\caption{\label{joint2} (color online). {\bf Joint complex-field PDF of speckles with distinct intensity PDFs at three planes}. All parameters are identical to those in Fig. 4. Speckle intensity PDFs at $z = 0, d, 2d$ are unimodal, flat, bimodal in (a); super-Rayleigh, sub-Rayleigh, super-Rayleigh in (b). Despite of axial-varying field-amplitude PDF, the phase PDF remains uniform over $[0, 2\pi)$. Joint complex-field PDFs are obtained by averaging over 30 speckle realizations.}
\end{figure*}

To confirm the volumetric speckles are fully developed, we compute their joint complex-field PDFs \cite{bender2019creating}. The phase of a customized speckle field is retrieved from the measured T-matrix. Circularity of the joint complex-field PDFs, shown in Fig. S\ref{joint1}, indicates the phase is uniformly distributed over [0, 2$\pi$).  Hence, the customized speckles are fully developed. 
\newline \indent For three-plane speckle customization in Fig. 4, the joint complex-field PDFs are shown in Fig. S\ref{joint2}. Despite the variation of the field-amplitude PDF, the phase PDF remains uniform across [0, 2$\pi$). Thus the amplitude and phase values of the speckle field are uncorrelated.  

\bibliography{supplement_arXiv}